\documentclass[
tightenlines,
showpacs
,showkeys
,nofootinbib
,aps
,amsfonts
,amssymb
]{revtex4-1}

\usepackage{graphicx}  
\usepackage{amsmath,amssymb,bm} 
\usepackage[dvips]{color}
\usepackage[normalem]{ulem}  
 \usepackage{pslatex}  
 \usepackage{hyperref}

\newcommand{\bce}{\begin{center}}  
\newcommand{\ece}{\end{center}}
\newcommand{\beq}{\begin{equation}}  
\newcommand{\eeq}{\end{equation}}
\newcommand{\beqy}{\begin{eqnarray}}
\newcommand{\eeqy}{\end{eqnarray}}

\def\r{{\it r}}

\def\){\right)} 
\def\({\left(} 
\def\]{\right]} 
\def\[{\left[}

\begin{document}

\title{Neutrino cooling and spin-down of rapidly rotating compact stars}

\author{%
 Prashanth Jaikumar$^{1,2}$, Stou Sandalski$^1$}

\affiliation{$^1$California State University Long Beach, Long Beach, CA 90840 USA}
\affiliation{$^2$Institute of Mathematical Sciences, C.I.T Campus, Chennai, 
TN 600113, India}

\begin{abstract} 

The gravitational-wave instability of \r-modes in rapidly rotating compact stars is  believed to spin them down to angular frequencies $\Omega\sim 0.1\Omega_{\rm Kepler}$ soon after their birth in a Supernova. We point out that the \r-mode perturbation also impacts the neutrino cooling and viscosity in hot compact stars via processes that restore weak equilibrium. We illustrate this fact with a simple model of spin-down due to gravitational wave emission in compact stars composed entirely of three-flavor degenerate quark matter (a strange quark star). Non-equilibrium neutrino cooling of this oscillating fluid matter is quantified. Our results imply that a consistent treatment of thermal and spin-frequency evolution of a young and hot compact star is a requisite in estimating the persistence of gravitational waves from such a source. 

\end{abstract}
 
\maketitle

\section{Introduction}
\label{sec_intro}

\begin{figure}{h!}
\label{fig1}
\centering
\includegraphics[scale=0.5,angle=270]{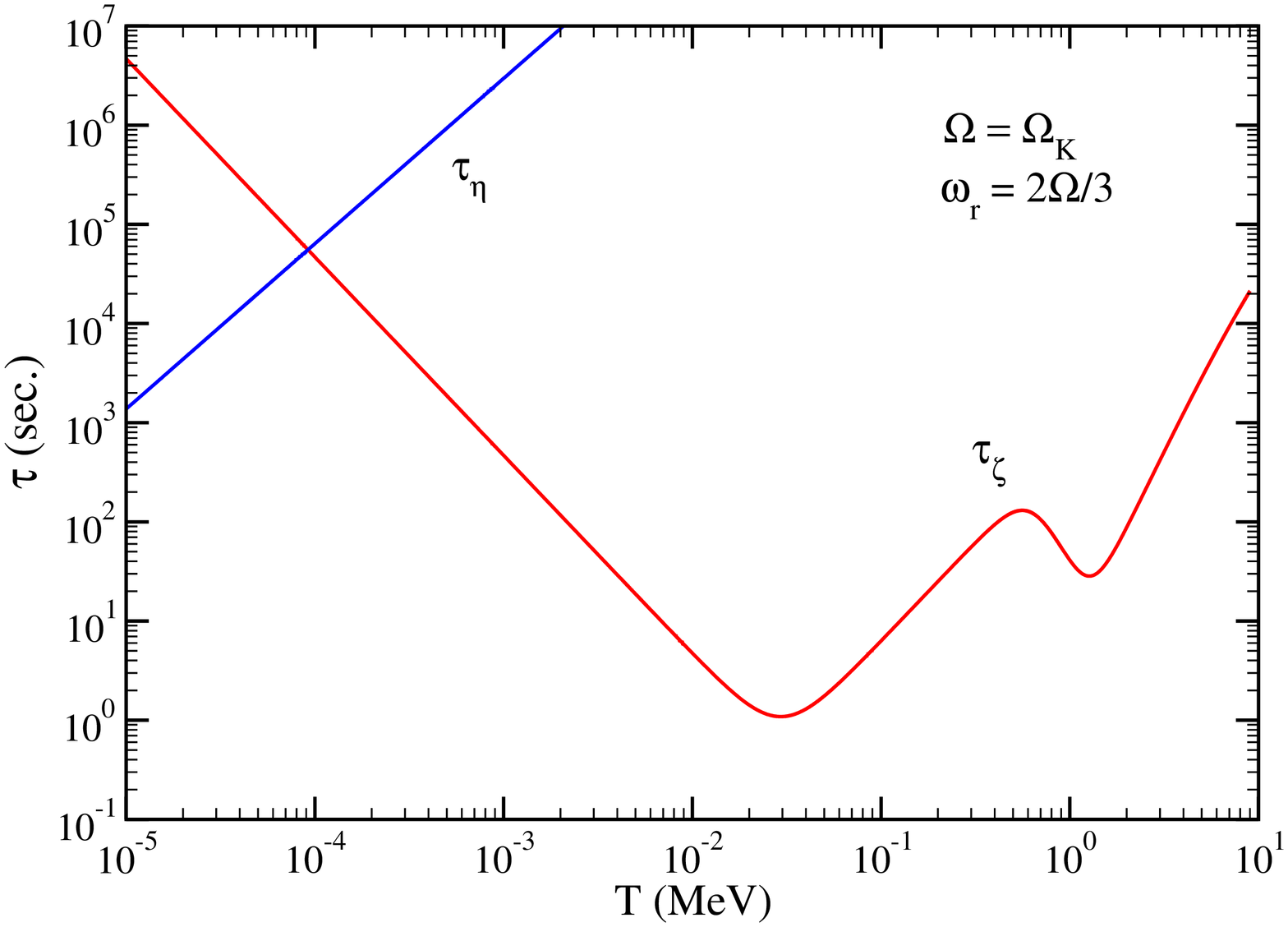}
\vskip 1.0cm
\includegraphics[scale=0.5,angle=270]{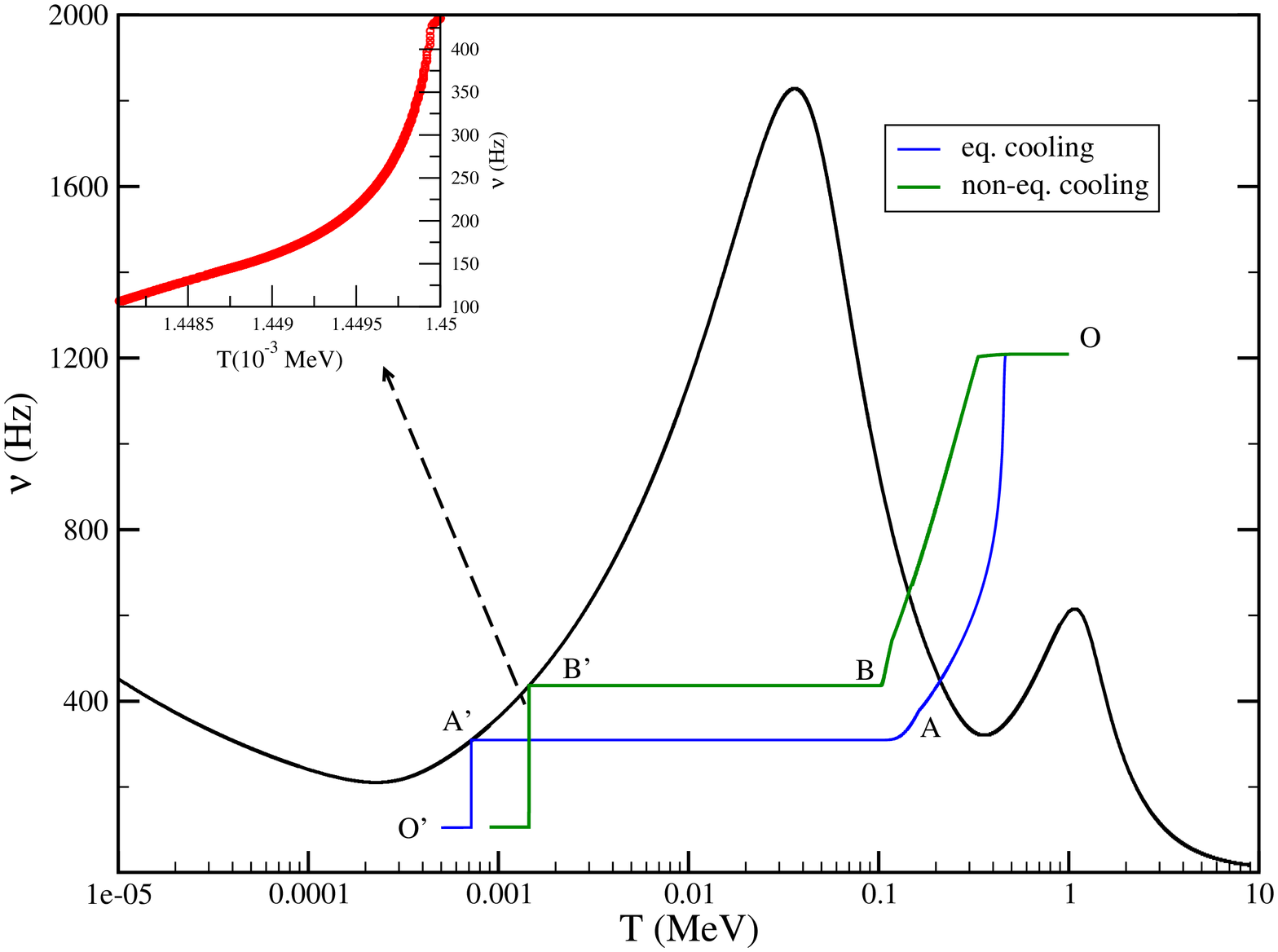}
\caption{{\it Top panel:} The temperature dependence of the shear ($\eta$) and bulk ($\zeta$) viscosity damping timescales 
from Eqs.~(\ref{tshearv}), (\ref{tbulkv}). The stellar gravitational mass and radius (Newtonian gravity) is 
1.49$M_{\odot}$ and 9.47 km respectively, while the r-mode frequency is 0.88 kHz from Eqn.~(\ref{modefreq}). {\it Bottom Panel:}
Spin down history versus temperature for equilibrium and non-equilibrium neutrino cooling (see text in Section~\ref{sec_evo} for details).}
\label{fig:visc}
\end{figure}

Neutron stars that undergo pulsations of quadrupole order or higher can emit gravitational waves. Such quasi-normal pulsation modes are typically expected in the aftermath of a supernova or a sudden rearrangement of the neutron star crust or core (egs., a starquarke or a hadron-quark phase transition). 
The possibility of detecting the associated gravitational radiation in current experiments such as LIGO and VIRGO is enhanced by the existence of the \r-mode instability in rotating neutron stars. Discovered by Andersson~\cite{Andersson:1997xt}, and detailed in subsequent works~\cite{Friedman:1997uh,Lindblom:1998wf}, \r-modes are fluid pulsations whose dynamics are governed by the Coriolis force, and the $l\geq 2$ modes grow by emitting gravitational waves in all rotating neutron stars irrespective of their rotation speed. So long as viscous damping effects are small, which is the case for hot young rapidly rotating neutron stars, the \r-mode can become unstable to the emission of gravitational waves while the star spins down by losing angular momentum (the Chandrasekhar-Friedman-Schutz or CFS instability~\cite{Chandra:1970,FS:1978}). This is one of several reasons why neutron stars are regarded as promising targets of opportunity for gravitational wave detection~\cite{Schutz:2010xm}.

\vskip 0.2cm

For experimental detection of gravitational waves (see eg.,~\cite{Schutz:2010xm}  for a pedagogical treatment), it is important to have  theoretical predictions for the gravitational waveform (eg. $\bar{h}(f)$ in the
frequency-domain). According to the simple model discussed in~\cite{Owen:1998xg} which we employ in this work, the angle-averaged signal in the time domain $h(t)$ depends on the distance, instantaneous angular velocity $\Omega(t)$ and average density of the star, as well as the \r-mode amplitude $\alpha(t)$. The time evolution of the mode amplitude and of the angular velocity constitute two coupled first-order differential equations which can be solved explicitly for a particular stellar configuration once viscous damping and gravitational wave timescales are known. Since the viscous damping timescale is very sensitive to the temperature of the fluid, thermal evolution of the star affects \r-mode evolution. In previous works~\cite{Owen:1998xg,Bildsten:1999zn,Bondarescu:2007jw}, while this fact has been recognized, the converse effect, viz., that the \r-mode perturbation also affects the cooling rate of the neutron star has been missed. A typical \r-mode perturbation induces chemically equilibrating weak reactions that alter the neutrino emissivity from the standard equilibrium expressions and hence impact the cooling rate quantitatively. Therefore, adopting a simple power law for the cooling rate based on the temperature-dependence of equilibrium emissivities may not accurately describe the spin-down of the star. Thermal evolution, \r-mode evolution and spin-down (i.e, rotational evolution) of a young and hot compact star are all coupled and a consistent analysis of these must be performed in any study of persistent gravitational waveforms from a rapidly rotating compact star. 

\vskip 0.2cm

In this paper, we illustrate this generic interplay of thermochemical and spin-down effects and quantify it in a particular case, viz., that of \r-modes in three-flavor ($u,d,s$) degenerate quark matter. Such a phase can arise in the ultra-dense interior of a neutron star or comprise almost the entire matter in a quark star~\footnote{Although cold and dense quark matter is believed to be in a color superconducting state\cite{Alford:1998mk}, we will not consider additional facets introduced by Cooper pairing here, leaving such considerations to future work. For \r-modes in color superconducting quark matter, see~\cite{Jaikumar:2008kh,Sa'd:2008gf,Andersson:2010sh,Rupak:2010qg}}. We also take into account leptonic (Urca) contributions to the viscosity of strange quark matter, calculated recently~\cite{Sa'd:2007ud}. Our main findings are: (i) the star can undergo two distinct epochs of gravitational wave emission and rapid spin-down, of very different duration, separated by a decades-long hiatus where its rotation rate is stable;
(ii) the leptonic viscosity can be important for the spin-down history of the star if the initial spin-frequency $\nu\lesssim 600$ Hz ($\nu\approx 0.5\Omega_{\rm Kepler}/(2\pi)$); (iii) the star takes about an order of magnitude shorter time to spin down to a stable frequency if non-equilibrium neutrino emissivities are considered and the evolution equations are evolved consistently.

\vskip 0.2cm

In Section \ref{sec_mode}, we provide a general overview of the link between \r-modes, viscosity and gravitational waves, outlining the phenomenological model for mode and spin evolution introduced in~\cite{Owen:1998xg} and adopted for this work. In Section \ref{sec_neutrino}, we explain how \r-mode perturbations affect the viscosity and neutrino emissivity of the constituent matter. In Section \ref{sec_evo}, we present results from a coupled analysis of thermal and spin-down evolution of a strange quark star. Section \ref{sec_conc} gathers our conclusions and discusses the implications for persistent gravitational waves from neutron/quark stars. We include some mathematical details in the Appendices.   

\section{\r-mode evolution and gravitational waves}
\label{sec_mode}
Pulsation modes of compact stars are classified by the nature of the restoring force in the perturbed Euler equation. Analysis in the co-rotating frame shows that the Coriolis force provides the restoring force for small non-radial fluid perturbations~\cite{Andersson:2000mf}. It is conventional to assume that purely toroidal mode solutions to the perturbed Euler equation exist even for rotating stars, in which case the velocity perturbation can be expressed to ${\mathcal {O}}(\Omega)$ as~\cite{Lindblom:1998wf}

\beq
\delta\vec{v}=R\Omega f(r/R)\vec{Y}_{lm}^B{\rm e}^{i\omega_{\rm rot}^{(l)} t}\,\,,
\eeq

with $R$ and $\Omega$ being the star's radius and angular velocity, $f$ and $\vec{Y}_{lm}^B$ the radial and angular functional dependence of the mode, and $\omega_{\rm rot}^{(l)}$ the mode frequency in the co-rotating frame. If we restrict ourselves to isentropic perturbations (no composition or temperature gradients), it can be shown that only the $l=m$ modes survive~\cite{Provost:1981}, that $f(r/R)=\alpha(r/R)^l$ with $\alpha$ an arbitrary constant (dimensionless mode amplitude), and that~\cite{PP:1978} 

\beq
\label{modefreq}
\omega_{\rm rot}^{(m)}=\frac{2\Omega}{(m+1)}+{\cal O}(\Omega^3)\,\,.
\eeq

These are the \r-modes and they are termed quasi-normal when they lose energy to gravitational waves ($l\geq 2$). The energy of the \r-mode is dissipated according to~\cite{Lindblom:1998wf}:

\begin{align}
\label{rmodeloss}
\frac{dE}{dt}=-(\omega_{\rm rot}-m\Omega)^{2m+1}\omega_{\rm rot}|\delta
J_{mm}|^2
-\int d^3r (2\eta\delta\sigma^{ab}\delta\sigma^*_{ab}+
\zeta\delta\sigma\delta\sigma^*) \,\,, 
\end{align}

where $\delta\sigma=\nabla_a\delta v^a$ is the volume
expansion due to the \r-mode and $\delta\sigma_{ab}=\frac{1}{2}
(\nabla_a\delta v_b+\nabla_b\delta
v_a-\frac{2}{3}\delta_{ab}\nabla_c\delta v^c)$ is (the traceless part of)
the shear tensor. The first term is the energy radiated in
gravitational waves to lowest order in $\Omega$ with $\delta J_{mm}$ being
the current multipole

\begin{align}  
\label{multipole}
\delta J_{mm}\propto
R^2\Omega\int_0^R dr \rho \left(\frac{r^2}{R}\right)^{(m+1)} \,\,. 
\end{align}

For $m\geq 2$, $\omega_{\rm rot}<m\Omega$, 
so that the $r$-mode energy
grows with gravitational wave emission, triggering the \r-mode instability. 
 Viscosity from a variety of mechanisms in the fluid can suppress $r$-mode growth, and counter the instability. The fate of the \r-mode is determined by a competition between the timescale for viscous damping and gravitational wave emission. According to the simple model for mode growth in the linear regime described in~\cite{Owen:1998xg}, the evolution equations for $\alpha$ (mode amplitude) and $\Omega$ (angular velocity) are:

\beqy
\label{evoeqs}
\dot{\Omega}&=&-\frac{2\Omega}{\tau_V}\frac{\alpha^2Q}{1+\alpha^2Q}\,\,,\\ \nonumber
\dot{\alpha}&=&-\frac{\alpha}{\tau_{GR}}-\frac{\alpha}{\tau_V}\frac{1-\alpha^2Q}{1+\alpha^2Q}\,\,,
\eeqy

where the viscous damping timescale is $\tau_V={(\tau_{\zeta}^{-1}+\tau_{\eta}^{-1})}^{-1}$ ($\zeta$: bulk viscosity, $\eta$: shear viscosity) and the gravitational wave growth timescale is $\tau_{GR}$. In general, $\tau_i^{-1}=-(dE_i/dt)/2E;\,i=\zeta,\eta,GR$. Reading off $(dE_i/dt)$ from Eqn.~(\ref{rmodeloss}), and using the lowest order expression
for the \r-mode energy $E$, one obtains the expressions for the damping timescales $\tau$ provided in Appendix B. The quantity $Q=3\bar{J}/2\bar{I}$ where~\cite{Owen:1998xg}

\beq
\bar{J}=\frac{1}{MR^4}\int_0^R\rho r^6 dr;\quad \bar{I}=\frac{8\pi}{3MR^2}\int_0^R \rho r^4 dr \,\,,
\eeq

with $M$ being the stellar gravitational mass and $\rho$ the energy density. In this model~\cite{Owen:1998xg}, exponential mode growth is followed by non-linear saturation characterized by $\alpha^2\sim\alpha_0^2\sim 1$ where $\dot{\alpha}\approx 0$. While such large saturation amplitudes are probably unrealistic due to multi-mode coupling and onset of non-linear effects~\cite{Arras:2002dw,Bondarescu:2007jw,Bondarescu:2008qx}, we prefer to use this large value to demonstrate the qualitative effect induced by the coupling of temperature and mode evolution. Ultimately, our analysis should  be applied to more realistic saturation amplitudes and include the coupling to
daughter modes~\cite{Bondarescu:2007jw,Bondarescu:2008qx}. Within our working model~\cite{Owen:1998xg}, once saturation is achieved, only the angular velocity evolves

\beq
\label{evoeqs2}
\dot{\Omega}=\frac{2\Omega}{\tau_{GR}}\frac{\alpha_0^2Q}{1-\alpha_0^2Q}\,\,.
\eeq

It is important to note that $\tau_V$ is strongly temperature dependent while $\tau_{GR}$ is temperature-independent so that solution of Eqn.~(\ref{evoeqs}) requires an estimate of the thermal evolution $T(t)$, which is determined in the simplest case (ignoring thermal gradients and conduction) by

\beq
\label{cooling}
C_v\frac{dT}{dt}=-L_{\nu}+L_{h}\,\,,
\eeq

where $C_v$ is the specific heat, $L_{\nu}$ is the neutrino luminosity and $L_{h}$ is the heating rate due to dissipation~\footnote{Viscous heating by shear viscosity, estimated as in~\cite{Owen:1998xg}, has a small effect on the temperature evolution in case of strange quark matter, while local heating due to bulk viscosity is rapidly radiated away in neutrinos and is unimportant for even the smallest timescales required for evolving eqs~(\ref{evoeqs}). Therefore, we only include viscous heating from shear viscosity in $L_h$.}. While the \r-mode is growing, the local baryonic number density fluctuates according to

\beq
n_B(t)=n_B^0+\delta n_B(t)=n_B^0+\delta\bar{n}_B{\rm Re}({\rm e}^{i\omega t})\,\,,
\eeq

where $\omega\equiv\omega_{\rm rot}^{(2)}$ and $\delta\bar{n}_B$ (proportional to $\alpha$, see Eqn.~(\ref{densitypert})) is the magnitude of 
the fluctuation. These fluctuations, which first appear at ${\cal O}(\Omega^2)$ rather than ${\cal O}(\Omega)$, destroy weak equilibrium, considerably enhancing the neutrino emissivity above its equilibrium value. They also affect the bulk viscosity, so that $\zeta$ and $L_{\nu}$ must be evaluated in the fluctuating foreground created by the \r-mode. In the next section, we describe this computation, highlighting the effect of the \r-mode perturbation on $\zeta$ and $L_{\nu}$.

\section{\r-modes, Viscosity \& Non-equilibrium neutrino rates}
\label{sec_neutrino}

The \r-mode perturbation described by the density fluctuation Eqn.~(\ref{densitypert})
causes a change in the fluid volume as well as the chemical potentials
of the electron ($e$) and the up ($u$), down ($d$) and strange ($s$) quarks.
Since weak equilibrating processes are relatively slow, 
the pressure fluctuation, expressed as $\delta P=\delta P_0{\rm e}^{i\omega t}$, lags the density fluctuation through the
former's dependence on the fluctuating chemical potentials (i.e., $\delta P_0$ acquires a phase). The dependence
on temperature fluctuations is ignored since, as mentioned before, dissipated heat can be carried 
away much faster than weak equilibration timescales.
Both leptonic and non-leptonic processes in dense quark matter can contribute to the viscosity at temperatures $T\geq 10^{10}$K for typical \r-mode frequencies $\omega\sim 1$kHz~\cite{Sa'd:2007ud}. Following the approach of~\cite{Sa'd:2007ud}, it is useful to choose the two independent fluctuations in the chemical potentials corresponding to the leptonic processes as $\delta\mu_1=\delta(\mu_d-\mu_u-\mu_e)$ and $\delta\mu_2=\delta
(\mu_s-\mu_u-\mu_e)$ which can be expressed in terms of
variations in the following quantities: baryon number ($\delta n_B/n_B$), charge ($\delta X_e$) and strangeness ($\delta X_s$)

\beq
\delta\mu_i=A_i\delta X_s+B_i\delta X_e+C_i\left(\frac{\delta n_B}{n_B}\right);
\quad X_s=\frac{n_s}{n_B}~\,,\quad X_e=\frac{n_e}{n_B}\,\,,
\eeq

where $n_s$ and $n_e$ are the strange quark and electron number densities.
The non-leptonic process $d+u\leftrightarrow u+s$ is characterized by
$\delta\mu_3=\delta\mu_1-\delta\mu_2$.
The bulk viscosity, which is a measure of the power dissipated per 
unit volume over an oscillation cycle is given by 

\beq
\zeta=-\frac{1}{\omega}{\rm Im}(\delta P_0)\frac{n_B^0}{\delta\bar{n}_B};\quad {\rm Im}(\delta P_0)=-n_B\left[C_1{\rm Im}(\delta X_{e,0})+(C_2-C_1){\rm Im}(\delta X_{s,0})\right]\,\,.
\eeq
 
 For the density perturbation, we adopt an expression derived in~\cite{Lindblom:1998ka} (also Eqn.~(6.8) of~\cite{Lindblom:1999yk})

\beqy
\label{densitypert}
\frac{\delta\bar{n}_B}{n_B}&=&R^2\Omega^2\frac{d\rho}{dP}\left[\delta U_0+\delta \Phi_0\right]\,\,,\\ \nonumber
\delta U_0&=&\alpha\left(\frac{r}{R}\right)^{m+1}P_{m+1}^m({\rm cos}~\theta){\rm e}^{im\phi};\quad \delta \Phi_0=\alpha\delta\phi_0(R)P_{m+1}^m({\rm cos}~\theta){\rm e}^{im\phi}\,\,,
\eeqy

where $\delta\Phi_0$ is the change in gravitational potential due to the \r-mode oscillation ($\delta\phi_0(R)$ is obtained from Eq.(4.4) of~\cite{Lindblom:1999yk}). The perturbation, and hence $\delta\mu_i$, has a radial and angular dependence, leading to different emissivities and viscosities in different directions and radial distances. We take a much simplified approach whereby, as is common in interpretation of neutron star thermal observations, the temperature $T$ refers to or is related to a {\it local} surface temperature map (eg., a "hot spot")~\cite{Page:1994dz}. Even if full angular dependence is retained, additional anisotropy in the bulk viscosity can be induced by strong magnetic fields in quark matter~\cite{Huang:2009ue}. For now, we proceed with our simplifying assumptions of homogeneity and isotropy with magnetic fields set to zero. These assumptions should be relaxed in fully 3-dimensional calculations, which require extensive computations on a spatial grid. The computation of ${\rm Im}(\delta X_{e,o}), {\rm Im}(\delta X_{s,0})$ and the co-efficients $A_i,B_i,C_i$ is straightforward once an equation of state is specified (see Appendix A for the equation of state for degenerate strange quark matter adopted in this work). We find

\beqy
\label{zeta}
\zeta&=&\frac{n_B}{\omega}\frac{1}{g_1^2+g_2^2}\sum_{(i,j)=1;i<j}^3\alpha_i\alpha_j\theta_{ij}\,\,,\\ \nonumber
g_1&=&(A_1B_2-A_2B_1)\Sigma-\Pi;\,\,\,g_2=\alpha_1\alpha_2(A_2-A_1)+\alpha_1\alpha_3(A_2-B_2)-\alpha_2\alpha_3B_1\,\,,\\ \nonumber
\Sigma&=&\alpha_1+\alpha_2+\alpha_3;\,\Pi=\alpha_1\alpha_2\alpha_3\,\,,\nonumber\\
\Theta_{12}&=&\Sigma(A_1C_2-A_2C_1)^2+\Pi(C_1-C_2)^2\,\,,\,\Theta_{13}=\Sigma\left[C_1(A_2-B_2)^2+C_2B_2\right]^2+\Pi C_2^2\,\,,\,\Theta_{23}=\Sigma(C_1B_2-C_2B_1)^2+\Pi C_2^2\,\,,
\eeqy

where $A_i,B_i,C_i$ are given in Appendix A. In Eqn.~(\ref{zeta}), $\alpha_i=n_B\omega/\lambda_i$ where the $\lambda_i$s are obtained from 
the decay rates ($\Gamma$) of the non-equilibrium processes at first order as

\beqy
\label{decay}
\Gamma_{\rm net}^{(1)}&=&\Gamma_{d\rightarrow u+e^-+\bar{\nu}_e}-\Gamma_{u+e^-\rightarrow d+\nu_e}\approx\lambda_1\delta\mu_1\,\,,\\ \nonumber \Gamma_{\rm net}^{(2)}&=&\Gamma_{s\rightarrow u+e^-+\bar{\nu}_e}-\Gamma_{u+e^-\rightarrow s+\nu_e}\approx\lambda_2\delta\mu_2\,\,,\\ \nonumber \Gamma_{\rm net}^{(3)}&=&\Gamma_{u+d\rightarrow s+u}-\Gamma_{s+u\rightarrow u+d}\approx\lambda_3(\delta\mu_1-\delta\mu_2)=\lambda_3\delta\mu_3\,\,,
\eeqy

where the second relation in each line is a commonly made approximation to first order in $\delta\mu$. The net decay rates for the leptonic processes can be computed exactly from the corresponding non-equilibrium neutrino rates~\cite{FloresTulian:2006fq}

\beq
\label{decayemiss}
\frac{\partial\epsilon_{\rm tot}^{(i)}}{\partial\delta\mu_i}=3(\Gamma_{f}^{(i)}-\Gamma_{b}^{(i)})=3\Gamma_{\rm net}^{(i)};\quad i=1,2
\eeq

where $\Gamma_f$ and $\Gamma_b$ are the forward and backward reaction rates respectively, and $\epsilon_{\rm tot}^{(i)}=\epsilon_{f}^{(i)}+\epsilon_{b}^{(i)}=\epsilon_{\bar{\nu}_e}^{(i)}+\epsilon_{\nu_e}^{(i)}$. As shown in Appendix B, the relevant non-equilibrium neutrino rates $\epsilon_{\rm tot}^{(i)}$ are simply related to the equilibrium emissivities ($\epsilon_{\rm tot}^{(i),{\rm eq}}$) as~\cite{Reise:1995}

\beq
\label{noneqemiss}
\epsilon_{\rm tot}^{(i)}=\epsilon_{\rm tot}^{(i),{\rm eq}}\left[1+\frac{1071u_i^2+315u_i^4+21u_i^6}{457}\right];\quad u_i=\frac{\delta\mu_i}{\pi kT}\,\,.
\eeq

Combining Eqns.(\ref{decay}),(\ref{decayemiss}) and (\ref{noneqemiss}) with Eqn.~(\ref{deltamueqn}) generates a self-consistent solution for the $\delta\mu_i$ even for large $u_i$. Since it turns out that $\delta\mu_i\gg kT$ can occur for the r-mode perturbation, we retain the full
expression in Eqn.~(\ref{noneqemiss}). Inclusion of these "large-amplitude" rates has been recently emphasized in the context of \r-modes~\cite{Alford:2010gw}. The decay rate for the non-leptonic process $\Gamma_{\rm net}^{(3)}$ has been calculated in various works. We adopt the analytic result of Madsen~\cite{Madsen:1992sx,Madsen:1993xx}, which improves Sawyer's result~\cite{Sawyer:1989} by including a $\delta\mu_3^3$ term as well as $m_s\neq 0$ corrections (their inclusion increases the rate). This approximation matches quite well with the exact numerical result up to $T\sim 10^{11}$K, which is the upper temperature bound considered in this work.
The equilibrium neutrino emissivities appearing in Eqn.~(\ref{noneqemiss}) are
~\cite{Iwamoto}

\beqy
\epsilon_{\rm tot}^{(1),{\rm eq}}&=&\frac{457}{630}\alpha_sT^6G_F^2{\rm cos}^2~\theta_c\mu_e\mu_d\mu_u\,\,, \\ \nonumber
\epsilon_{\rm tot}^{(2),{\rm eq}}&=&\frac{457\pi}{1680}T^6G_F^2{\rm sin}^2~\theta_c\mu_sm_s^2\,\,.\\ \nonumber
\eeqy

\section{Neutron star cooling and spin-down}
\label{sec_evo}
 As mentioned in the previous section, $\delta\mu_i$ can be much larger than $kT$ over some time domain of the evolution. In our simulation, this happens in the initial few hundreds of seconds of the star's spin down due to the \r-mode instability. Consequently, non-equilibrium neutrino emissivities are much higher than equilibrium values, and we should anticipate a change in the cooling history of the star.
In Fig~\ref{fig1} (bottom panel), we display results for the angular velocity versus temperature evolution from the combined analysis of Eqn.(\ref{evoeqs}), for which we used the bulk and shear viscosity damping timescales displayed in the top panel of the same figure. Cooling with equilibrium and non-equilibrium neutrino emissivities (Eqn.~(\ref{noneqemiss}))
is studied. 

Firstly, we note that the double-valley feature of the damping timescale (top panel) is reflected in the double-peak feature of the critical frequency curve, which is the locus of points satisfying $1/\tau_{\rm f}(T)|_{\Omega_c}=[1/\tau_{\zeta}(T)+1/\tau_{\eta}(T)+1/\tau_{GW}(T))] _{\Omega_c}=0$. The critical frequency curve marks the boundary between rapid spin-down and unstable mode growth accompanied by gravitational waves in the region above the curve and stable rotation with a decaying \r-mode amplitude below it. As expected, the critical frequency curve mirrors the behaviour of the bulk viscosity of strange quark matter (see eg., Fig. 3 of~\cite{Sa'd:2007ud}). The larger of the two peaks corresponds to the maximum in the bulk viscosity of the non-leptonic process~\cite{Madsen:1992sx} while the smaller peak is from the maximum in the leptonic process~\cite{Sa'd:2007ud}. The peak height is inversely proportional to the \r-mode frequency ($\omega=0.88$kHz for this figure). For the simple model assumed here, these features determine the evolution of the angular velocity with temperature, as the star cools and spins down. 

To determine the persistence of gravitational waves from \r-modes in a cooling compact star, we track the angular momentum evolution against the critical frequency curve to see when the star enters/exits the stable region. In the
case of equilibrium neutrino rates, which ignores the effect of the \r-mode 
on neutrino emissivities, we find that the star rapidly spins down in about 
a few hours ($\approx 10^4$ seconds) from its initial spin frequency, assumed to be the Kepler frequency ($\Omega_K/(2\pi)\approx 1200$ Hz) down to about a third of the starting value. This is the segment $OA$ in Fig.~\ref{fig1} (bottom panel). The time to spin down to a stable rotation rate does not depend on the initial mode amplitude (typically, we varied this between $10^{-6}$ and $10^{-3}$), nor does it depend on the mode saturation amplitude $\alpha_{0}$, which we varied between $0.3-1.0$. However, the stable frequency at which the star settles down upon entering the stable region depends on $\alpha_{0}$, becoming smaller with increasing $\alpha_0$. Once the star enters this stable region, where bulk viscosity is large enough to damp the \r-mode, the mode amplitude decreases to negligible values. Since the \r-mode is damped, it does not affect the neutrino rates in any way and equilibrium rates apply. 
While a small time step of about 0.01 seconds is required in the initial rapid cooling phase, the time step in the stable region can be safely increased to
hundreds of seconds, since temperature evolution and angular velocity evolution 
is extremely slow. The star takes of order $10^9$ seconds (nearly 100 years) to cool down to the point where it can once again enter the instability region. This hiatus between spin-down epochs is depicted by the segment $AA^{\prime}$.
In the interim, the star has cooled by two orders of magnitude down to $T\approx 0.7$ keV, but hardly changed its spin-frequency (here, we do not take into account spin-down due to dipole braking or the possibility of glitches). Upon entering the unstable region at $A^{\prime}$, the mode amplitude begins a rapid growth from a negligibly small value and saturates within a year. At this point, the star spins down again due to the CFS instability and takes ${\cal O}(10^8)$ seconds (10 years) to reach a stable value of $\Omega/(2\pi)\sim 100$ Hz. During this phase, the star is expected to once again radiate gravitational waves. Thus, a strange quark star might be expected to become unstable to \r-modes over two distinct epochs, separated by about a 100 years, with the second one lasting
about 10,000 times longer than the first. 

For a consistent and more accurate estimate of the persistence of gravitational waves, non-equilibrium neutrino rates must be taken into account. A clear difference can be seen in the angular momentum history in Fig.~\ref{fig1}. 
Although we begin with the same initial conditions, neutrino cooling is more rapid, since the emissivities are larger (Eqn.~(\ref{noneqemiss})). Consequently, the star enters the stable region at a higher angular velocity and in less time than the equilibrium case (segment $OB$). We find this time to be of order $10^3$ seconds (less than 1 hour). The cooling in the stable region is once again marked by slow cooling and a negligible \r-mode amplitude, resulting in equilibrium-type cooling. However, the fact that the star is at a higher stable rotation rate now implies that it spends less time in the stable region (segment $BB^{\prime}$). This lasts approximately $5.10^8$ seconds ($\approx$ 15 years), to be contrasted with the equilibrium case, where this era lasts for about 100 years. Upon exiting this region, the star spins down rapidly once more, taking $\approx 10^7$ seconds (4 months) to reach its final stable value. This is the segment $B^{\prime}O^{\prime}$, shown in enlarged inset view in Fig.~\ref{fig1}. The final rotation rate is exactly the same irrespective of the cooling history. Therefore, the importance of incorporating non-equilibrium  emissivities is only evident in the first 100 years or so of the star's spin-down history. Another important difference is the shorter time for which the star woud be active in gravitational waves if non-equilibrium rates are used. This indicates that the epoch of gravitational waves is shorter than expected, narrowing the window of detection. Similar results are expected for the case of neutron stars on the basis of the generality of the evolution equations, even though neutron matter has very different viscosities and cooling rates. 

\section{Conclusions}
\label{sec_conc}
We have shown that the \r-mode perturbation affects the neutrino cooling and viscosity in hot compact stars via processes that restore weak equilibrium. Using a simplified model of spin-down due to gravitational wave emission in compact stars composed entirely of three-flavor degenerate quark matter, we find that non-equilibrium neutrino rates associated to the \r-mode fluctuation change
considerably the cooling history of the star, as well as the angular velocity evolution, resulting in shorter epochs of gravitational wave emission for isolated quark stars. We have taken into account a recent calculation of leptonic contributions to the viscosity of quark matter, but find that it can only play a role if the initial spin frequency is chosen sufficiently small ($\approx 600$Hz, see Fig.~\ref{fig1}). This is because of the smaller viscosity of the leptonic process, specially at \r-mode frequencies characteristic of very rapidly rotating stars. However, the role of the non-leptonic process is very prominent. It leads to two distinct rapid spin-down events in the star's evolution, caused by the CFS instability towards gravitational wave emission. Although we have focused on quark matter here as an exemplar, it should be clear that a similar exercise can be performed for any sufficiently simple phase of dense matter, including beta-stable neutron-rich matter. It is important for \r-mode saturation and damping studies that one includes the large-amplitude bulk viscosity results, as argued in~\cite{Alford:2010gw}, as well as realistic saturation amplitudes~\cite{Bondarescu:2007jw}. While analytic approximations for the bulk viscosity exist for a simple quark matter equation of state like the one adopted in this work, a full numerical solution is required for hadronic matter, where non-linearities in the chemical equilibration rates can take more complex forms. Furthermore, non-equilibrium rates~\cite{Haensel:1992} and updated viscosities~\cite{Benhar:2007yj} need to be taken into account in performing a similar calculation with hadronic equations of state. Despite these complications, one can predict that for isolated compact stars made up of neutron-rich matter, since there is no stable region between $\approx 10^7-10^{10}$K at rapid rotation rates~\cite{Owen:1998xg,Jaikumar:2008kh}, non-equilibrium cooling is probably important only over the first year or more of cooling. 

We also mention some caveats that follow from the simplifications made in this work. The lowest order density fluctuation (Eqn.~(\ref{densitypert})) enters at order $\Omega^2$, while the expressions for the r-mode frequency (Eqn.(\ref{modefreq})) and the current multipole (Eqn.~(\ref{multipole})) are ${\cal O}(\Omega)$. Higher-order effects change the mode frequency by less than 10\% (see eg., Table 1 in~\cite{Jaikumar:2008kh}) while Eqn...~(\ref{multipole}) suffices for gravitational radiation from the current multipole. A more detailed
discussion of higher-order rotation effects is given in~\cite{Lindblom:1999yk}.
Bulk viscosity timescales used in this work are approximate, and can be improved by replacing the Eulerian expression for the density perturbation, Eqn.~(\ref{densitypert}) by a Lagrangian one~\cite{Lindblom:1999yk}. This can change the bulk viscosity damping timescale by an order of magnitude, but the critical frequency curve is hardly modified because the bulk viscosity has a steep temperature dependence. Thus, the results presented in the bottom panel of Fig.~\ref{fig1} are accurate in this respect, even though the bulk viscosity timescale, as shown in the top panel is only approximate. 

Non-equilibrium cooling can also be important for accreting neutron stars, which are believed to be persistent sources of gravitational waves~\cite{Andersson:1998qs}, undergoing spin-cycles as they get spun-up by
accretion torques and spun-down repeatedly by the \r-mode instability. Larger duty-cycles are expected for exotic phases, such as quark matter or hyperonic matter~\cite{Nayyar:2005th}, providing the motivation to extend our work to low-mass X-ray binaries. Even for isolated strange quark stars, Andersson has shown~\cite{Andersson:2001ev} that if a strange star enters the instability window spinning near the break-up limit, it will emit bursts of gravitational waves as the spin evolution straddles the critical frequency curve. We plan to explore this interesting case as well as the spin-cycles of LMXBs including non-equilibrium cooling rates and improved quark matter viscosites. Finally, we note that LIGO data sets~\cite{Abadie:2010hv} can already be adapted to search for imprints of the \r-mode instability in gravitational waves~\cite{Owen:2010ng}. In anticipation of the improved sensitivity of advanced LIGO, it is important to continue to examine various factors affecting the \r-mode evolution.

\section*{Acknowledgments}

We thank C. D. Roberts, T. Kl\"ahn, B. El-Bennich and R. Young for useful discussions. This work was supported in part by the U.S. Department of Energy, Office of Nuclear Physics, contract no. DE-AC02-06CH11357 and by the United States ARMY High Performance Computing Research Center's Research and Infrastructure Development Program.

\section*{Appendix A}
\label{appA}
With reference to Eqn.~(\ref{zeta}), the definitions 

\beqy
e_1&=&\Sigma(B_1C_2-B_2C_1);\quad e_2=(\alpha_1\alpha_2C_1-\alpha_1(\alpha_2+\alpha_3)C_2)\,\,,\\ \nonumber
f_1&=&\Sigma(C_1A_2-A_1C_2);\quad f_2=(\alpha_1\alpha_3C_2+\alpha_2\alpha_3C_1)\,\,,\\ \nonumber
g_1&=&\Sigma(A_1B_2-B_1A_2)-\Pi;\quad g_2=\alpha_1\alpha_2(A_2-A_1)+\alpha_1\alpha_3(A_2-B_2)C_1-\alpha_2\alpha_3B_1\,\,,
\eeqy

lead to the expressions for $\delta X_{s,0}$ and $\delta X_{e,0}$ 

\beq
\delta X_{s,0}=\frac{e_1+ie_2}{g_1+ig_2}\left(\frac{\delta\bar{n}_B}{n_B}\right);\quad
\delta X_{e,0}=\frac{f_1+if_2}{g_1+ig_2}\left(\frac{\delta\bar{n}_B}{n_B}\right)\,\,,
\eeq

from which it follows that

\beq
\label{deltamueqn}
\delta\mu_i(t)=\left(\frac{\delta\bar{n}_B}{n_B}\right)\left[C_i{\rm cos}~(\omega t)+\sqrt{\frac{e_1^2+e_2^2}{g_1^2+g_2^2}}A_i{\rm cos}~(\omega t-\phi_e)+\sqrt{\frac{f_1^2+f_2^2}{g_1^2+g_2^2}}B_i{\rm cos}~(\omega t-\phi_f)\right]
\eeq

with

\beq
{\rm tan}\phi_e=\frac{e_1g_2-e_2g_1}{e_1g_1+e_2g_2};\quad {\rm tan}\phi_f=\frac{f_1g_2-f_2g_1}{f_1g_1+f_2g_2}\,\,.
\eeq

Given the following thermodynamic potential for 3-flavor degenerate quark matter (with electrons)

\beqy
\Omega&=&\sum_{u,d,s,e}\Omega_i;\quad \Omega_e=-\frac{\mu_e^4}{12\pi^2}\,\,,\\ \nonumber
\Omega_f&=&-\frac{\mu_f^4}{4\pi^2}\left\{\sqrt{1-z_f^2}\left(1-\frac{5}{2}z_f^2\right)+\frac{3}{2}z_f^4{\rm ln}\left[F(z)\right]-\frac{2\alpha_s}{\pi}\left[3\left(z_f^2{\rm ln}\left[F(z)\right]-\sqrt{1-z_f^2}\right)^2-2(1-z_f^2)^2\right]\right\} \,\,,\\ \nonumber
F(z)&=&\frac{1+\sqrt{1-z_f^2}}{z_f};\quad z_f=m_f/\mu_f;\quad \alpha_s=g_s^2/4\pi\,\,,
\eeqy

where $f$ denotes flavor, the coefficients $A_i ,B_i, C_i$ follow from

\beqy
A_1&=&-n_B\frac{\partial\mu_d}{\partial n_d};\quad B_1=-n_B\left[\frac{\partial\mu_d}{\partial n_d}+\frac{\partial\mu_u}{\partial n_u}+\frac{\partial\mu_e}{\partial n_e}\right];\quad C_1=\left[n_s\frac{\partial\mu_s}{\partial n_s}-n_d\frac{\partial\mu_d}{\partial n_d}\right] \,\,,\\ \nonumber
A_2&=&n_B\frac{\partial\mu_s}{\partial n_s};\quad B_2=-n_B\left[\frac{\partial\mu_u}{\partial n_u}+\frac{\partial\mu_e}{\partial n_e}\right];\quad C_2=\left[n_s\frac{\partial\mu_s}{\partial n_s}-n_u\frac{\partial\mu_u}{\partial n_u}-n_e\frac{\partial\mu_e}{\partial n_e}\right] \,\,,
\eeqy

upon using $n_i=-\partial\Omega_i/\partial\mu_i$ and the expressions

\beq
n_f\frac{\partial\mu_f}{\partial n_f}=\frac{\mu_f}{3}\left\{1-z_f^2+\frac{4\alpha_s}{\pi}z_f^2\left[{\rm ln}(2z_f^{-1})-\frac{2}{3}\right]\right\};\quad
n_e\frac{\partial\mu_e}{\partial n_e}=\frac{\mu_e}{3}\,\,.
\eeq

\section*{Appendix B}
\label{appB}
{\it Neutrino emissivity}: Omitting volume normalizations which ultimately cancel, the emissivity is given by

\beq
\label{emissmain}
\epsilon_{\bar{\nu}_e} = \int \biggl( \, \prod_{i}
\frac{d^3 {\bm p}_i}{2E_i(2\pi)^3} \biggr) \,
\frac{d^3 {\bm k}}{2\omega_k(2\pi)^3} \,
\, (2 \pi)^4 \, \delta(E_f^{\rm tot}-E_i^{\rm tot}) \, \delta^3({\bm P}_f - {\bm P}_i) 
\, \biggl( \: \sum_{\text{spin}} \, 
\bigl| \, {\mathcal{M}} \bigr|^2 \biggr) 
\, \omega_{k} \, {\mathcal{F}}(E) ,
\eeq

where $({\bm p}_i, E_i)$ denote the momenta and energy of the incoming and 
outgoing quarks and $(\omega_k,{\bm k})$ label the neutrino energy and
momenta. The delta functions account for total energy and momentum 
conservation, and the function ${\cal{F}}(E)=\prod_{\rm in}f(E_i)
\prod_{\rm out}(1-f(E_i))$ is the product of occupation (incoming) and
blocking (outgoing) statistical factors for the quarks, with Fermi-Dirac 
distribution function $f(E_i)=(\exp(\beta E_i) + 1)^{-1}$ (the neutrinos
free stream). The matrix element ${\mathcal{M}}$ is calculated from the 
corresponding tree level Feynman diagrams (see~\cite{Sa'd:2007ud}) in the 
low-energy limit. It can be shown that for the non-equilibrium case 
($E_i\approx \mu_i+\delta\mu_i$), the neutrino energy integral in
the total emissivity expression is modified from the equilibrium case~\cite{Iwamoto} such that

\beq
\epsilon_{\bar{\nu}_e}+\epsilon_{\nu_e}\propto\int~d\epsilon_{\nu}\epsilon_{\nu}^3\left[J\left(\frac{\epsilon_{\nu}+\delta\mu}{kT}\right)+J\left(\frac{\epsilon_{\nu}-\delta\mu}{kT}\right)\right]; \quad J(x)=\frac{T^2}{2}\left[\frac{\pi^2+x^2}{1+{\rm e}^{x}}\right]
\eeq

yielding Eqn.~(\ref{noneqemiss}).

\vskip 0.2cm

{\it Damping timescales}: For viscous damping of the \r-mode, the shear viscosity damping timescale is 

\beq
\label{tshearv}
\frac{1}{\tau_{\eta}} = \frac{(m-1)(2 m+1)}{\int_0^R dr \rho r^{2m+2}}
\int_0^R dr \eta r^{2m} ,
\eeq

where the shear viscosity is taken from~\cite{Heiselberg:1993cr}, while the bulk viscosity damping timescale follows from 

\beqy
\label{tbulkv}
\frac{1}{\tau_{\zeta}}&=&-\frac{1}{2 E}\left(\frac{dE}{dt}\right)_{\zeta}\,\,,\\ \nonumber
E&=&\frac{\pi}{2 m} \left(m+1\right)^3 \left(2 m+1\right)!
R^4 \Omega^2 \int_0^R dr \rho \(\frac{r}{R}\)^{2m+2} \,\,.\\ \nonumber
\eeqy

The gravitational radiation growth time scale is~\cite{Friedman:1997uh}

\beq
\frac{1}{\tau_{GW}} = - \frac{32 \pi G \Omega^{2m+2}}{c^{2m+3}} 
\frac{\left(m-1\right)^{2m}}{\left[\left(2m+1\right)!!\right]^2} 
\left(\frac{m+2}{m+1}\right)^{2m+2} \int_0^R dr \rho r^{2m+2} \,\,.
\eeq

\end{document}